\documentclass[reprint,
 superscriptaddress,
 amsmath,amssymb,
 aps,
prl,
]{revtex4-2}

\usepackage[utf8]{inputenc} 
\usepackage[T1]{fontenc}    
\usepackage{hyperref}       
\usepackage{url}            
\usepackage{booktabs}       
\usepackage{amsfonts}       
\usepackage{nicefrac}       
\usepackage{microtype}      
\usepackage{graphicx}
\usepackage{xcolor, etoolbox}
\usepackage{amsmath,amsfonts,amsthm} 
\usepackage{mathtools}
\usepackage{ragged2e}
\usepackage{braket}      
\usepackage[english]{babel} 
\usepackage{enumitem}
\usepackage{xspace}
\usepackage{dsfont}
\usepackage{bm}
\usepackage{bbm}
\usepackage{dcolumn}
\usepackage{mathrsfs}
\usepackage{todonotes}

\usepackage{float}          

\newcommand{\Tr}{\textbf{Tr}}

\hypersetup{
    colorlinks,
    linkcolor=blue,
    citecolor=blue,
    urlcolor=blue,
    breaklinks=true 
}

\DeclarePairedDelimiter\abs{\lvert}{\rvert}

\begin{document}

\title{Area laws and thermalization from classical entropies in a Bose-Einstein condensate} 

\author{Yannick Deller}
    \affiliation{Kirchhoff-Institut f\"{u}r Physik, Universit\"{a}t Heidelberg, Im Neuenheimer Feld 227, 69120 Heidelberg, Germany}

\author{Martin G\"{a}rttner}
    \affiliation{Institut für Festkörpertheorie und Optik, Friedrich-Schiller-Universität Jena, Max-Wien-Platz 1, 07743 Jena, Germany}

\author{Tobias Haas}
    \email{tobias.haas@ulb.be}
    \affiliation{Centre for Quantum Information and Communication, École polytechnique de Bruxelles, CP 165, Université libre de Bruxelles, 1050 Brussels, Belgium}

\author{Markus K. Oberthaler}
    \affiliation{Kirchhoff-Institut f\"{u}r Physik, Universit\"{a}t Heidelberg, Im Neuenheimer Feld 227, 69120 Heidelberg, Germany}

\author{Moritz Reh}
    \affiliation{Kirchhoff-Institut f\"{u}r Physik, Universit\"{a}t Heidelberg, Im Neuenheimer Feld 227, 69120 Heidelberg, Germany} 
    \affiliation{Institut für Festkörpertheorie und Optik, Friedrich-Schiller-Universität Jena, Max-Wien-Platz 1, 07743 Jena, Germany}

\author{Helmut Strobel}
    \affiliation{Kirchhoff-Institut f\"{u}r Physik, Universit\"{a}t Heidelberg, Im Neuenheimer Feld 227, 69120 Heidelberg, Germany}    

\begin{abstract}
The scaling of local quantum entropies is of utmost interest for characterizing quantum fields, many-body systems, and gravity. Despite their importance, theoretically and experimentally accessing quantum entropies is challenging as they are nonlinear functionals of the underlying quantum state. Here, we show that suitably chosen classical entropies capture many features of their quantum analogs for an experimentally relevant setting. We describe the post-quench dynamics of a multi-well spin-1 Bose-Einstein condensate from an initial product state via measurement distributions of spin observables and estimate the corresponding entropies using the asymptotically unbiased $k$-nearest neighbor method. We observe the dynamical build-up of quantum correlations signaled by an area law, as well as local thermalization revealed by a transition to a volume law, both in regimes characterized by non-Gaussian distributions. We emphasize that all relevant features can be observed at small sample numbers without reconstructing the underlying state or measurement distributions, rendering our method directly applicable to a large variety of models and experimental platforms.
\end{abstract}

\maketitle

\textit{Introduction} --- 
The quantum entropy of a spatial subregion has proven to serve as a ubiquitous tool for studying the spatio-temporal structure of entanglement \cite{Horodecki2009} and its role in various quantum phenomena, including local thermalization \cite{Popescu2006,Abanin2019,Haas2020a,Haas2020b}, quantum phase transitions \cite{Osborne2002}, information scrambling \cite{Jozsa2003,Landsman2019,Xu2022} and black hole physics \cite{Bombelli1986,Srednicki1993,Callan1994,Solodukhin2011}. Arguably the most sought-after phenomenon in this context is the area law, which is signaled by a logarithmic growth of the local entropy for one-dimensional systems \cite{Calabrese2004,Calabrese2006,Plenio2007,Hastings2007,Amico2008,Calabrese2009,Casini2009,Peschel2009,Eisert2010}. It appears at short times after quenching the couplings of a locally interacting system, that was initially prepared in a product state \cite{Eisert2010,Calabrese2005,Calabrese2007,Nezhadhaghighi2014,Calabrese2016,Calabrese2020} -- a scenario that can be readily realized experimentally. At later times, the system typically thermalizes, and the local entropy instead obeys a volume law, allowing for a macroscopic description using only a few thermodynamic quantities like temperature.

The main drawback of quantum entropic descriptions for many-body phenomena is their reliance on the knowledge of the full density matrix, which grows exponentially with the number of microscopic constituents. This has so far restricted the experimental access of quantum entropies to systems consisting of a few particles \cite{Islam2015,Kaufman2016,Brydges2019}, as full tomography of the quantum state is, with no further assumptions, infeasible for larger systems approaching mesoscopic scales. For continuous systems, area laws have only been experimentally reported in a Gaussian scenario \cite{Tajik2023}, while generally applicable methods have remained elusive. 

Recently, the necessity of considering exclusively \textit{quantum} entropies to probe quantum phenomena has been questioned. Suitably chosen \textit{classical} entropies of (quasi-) probability distributions also encode area and volume laws \cite{Haas2023b}. This insight naturally overcomes the need to reconstruct the full quantum state, both for theoretical and experimental investigations. Thus, the observation of entropic scaling behavior becomes accessible for experimental platforms, which can directly sample from such distributions, see for example \cite{Leonhardt1995,Kirchmair2013,Haas2014,Strobel2014,Barontini2015,Wang2016,Kunkel2018,Kunkel2019,Kunkel2021}.

Here, we show that area and volume laws are observable in state-of-the-art experiments with multi-well spin-1 Bose-Einstein condensates (BECs) \cite{Hamley2012,Kawaguchi2012} by considering entropies of measurement distributions over spin observables. Starting from an initial product state, we find area laws being dynamically generated for intermediate evolution times following a quench, thereby confirming the growth of entanglement until the system thermalizes locally, where the same entropies exhibit volume law behavior. Importantly, we do so without making assumptions about the functional form of the state and only rely on observables that are directly obtainable in standard experimental readouts \cite{Strobel2014,Kunkel2018,Kunkel2019,Kunkel2021,Tajik2023} while reducing the sample complexity to a feasible level. We comprehensively discuss our modeling of the spinor BEC and the estimation of classical entropies, including systematic checks for validity and generality, in \cite{Haas2025}.

\textit{Notation} --- We use natural units $\hbar = k_{\text{B}} = 1$, write bold (normal) letters for quantum operators $\boldsymbol{O}$ (classical variables $O$) as well as their traces and equip vacuum expressions with an overbar, e.g. $\bar{\boldsymbol{\rho}}$.

\begin{figure*}[t!]
    \centering
    \includegraphics[width=.98\linewidth]{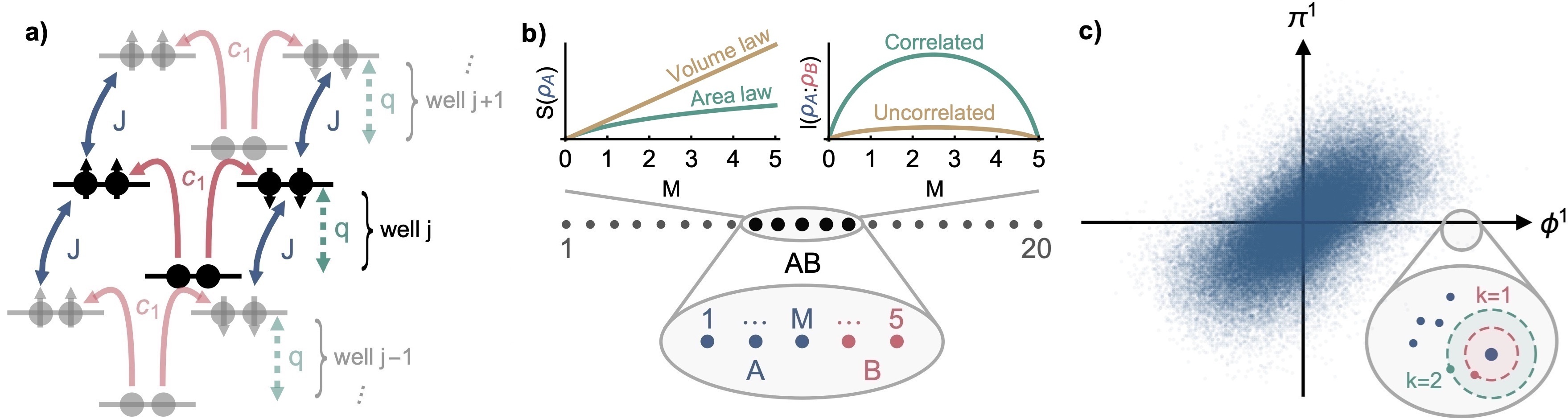}
    \caption{\textbf{a)} Illustration of relevant processes. The $\pm 1$ modes of each well are coupled to the 0 mode by spin-changing collisions with strength $c_1<0$ (red) and detuned by the quadratic Zeeman-shift $q>0$ (green). The atoms in the $\pm 1$ modes of well $j$ may hop to neighboring wells $j \pm 1$ with strength $J>0$ (blue). \textbf{b)} The full setup consists of 20 wells, from which we exclusively analyze the open system $AB$ given by the five wells $8-12$ (magnified inset). We partition this system into subsystems $A$ (blue) of size $M$ and $B$ (red) of size $5-M$ and study the scalings of information and correlation measures with $A$'s size $M$. \textbf{c)} Samples of the Wigner $W$-distribution of the left-most well in subsystem $A$ at time $t=4$. The entropy is estimated from samples using the $k$NN-estimator by analyzing the distribution of distances to the $k$-th.\ neighbor for each sample, see magnified inset for $k=1$ (red) and $k=2$ (green). Non-Gaussian features arise for higher-dimensional multi-well distributions, as measured by the relative entropy, see \cite{SM}.}
    \label{fig:Intro}
\end{figure*}

\textit{Multi-Well Spin-1 BEC} --- We consider a one-dimensional lattice of spin-1 BECs that extends over 20 wells, described by bosonic mode operators $[\boldsymbol{a}^j_{m_F},\boldsymbol{a}^{j' \dagger}_{m'_F}] = \delta^{j j'} \delta_{m_F m'_F}$ with $j \in \{1, ..., N \}$ and $m_F \in \{0, \pm 1 \}$. Starting from an initial product state with all zero modes ($m_F=0$) being occupied coherently with a mean number of $n=10^3$ atoms, we consider the evolution under the Hamiltonian
\begin{equation}
    \begin{split}
        \boldsymbol{H} &= \sum_{j=1}^{20} q \left( \boldsymbol{N}_1^{j} + \boldsymbol{N}_{-1}^{j} \right) + c_0 \, \boldsymbol{N}^j \left( \boldsymbol{N}^j - \mathds{1} \right) \\
        &+ c_1 \Big[ \left( \boldsymbol{N}^j_0 - (1/2) \mathds{1} \right) \left( \boldsymbol{N}^j_1 + \boldsymbol{N}^j_{-1} \right) \\
        &\hspace{1cm}+ \boldsymbol{a}_0^{j \dagger} \boldsymbol{a}_0^{j \dagger} \boldsymbol{a}_1^{j} \boldsymbol{a}_{-1}^{j} + \boldsymbol{a}_{1}^{j \dagger} \boldsymbol{a}_{-1}^{j \dagger} \boldsymbol{a}_0^{j} \boldsymbol{a}_{0}^{j} \Big] \\
        &-J\sum_{j=1}^{19}\sum_{m_F=\pm 1} \left( \boldsymbol{a}^{j\dagger}_{m_F}\boldsymbol{a}^{j+1}_{m_F}+\boldsymbol{a}^{(j+1)\dagger}_{m_F}\boldsymbol{a}^{j}_{m_F} \right).
    \end{split}
    \label{eq:FullHamiltonian}
\end{equation}
The single-well dynamics (first sum) includes density-density interactions $c_0 > 0$, the parameter $q>0$, which includes the quadratic Zeeman shift and is tunable via off-resonant microwave dressing, and spin-changing collisions $c_1 < 0$. Correlations among the wells build-up via nearest-neighbor interactions $J>0$ (second sum). We sketch all relevant contributions in  Fig.~\hyperref[fig:Intro]{1a)}, see \cite{Kawaguchi2012,Hamley2012,Stamper2013,Kunkel2018,Kunkel2019,Kunkel2021,Fujiwara2019,Jepsen2020,Huh2020,Kim2021,Kwon2022} for similar setups and \cite{Haas2025} for details.

For early times, the zero mode is occupied macroscopically, and the evolution is dominated by second-order fluctuations, such that \eqref{eq:FullHamiltonian} is well-approximated by an analytically solvable Gaussian model, which follows from treating the zero mode classically and dropping density-density interactions (see \cite{Haas2025} for details)
\begin{equation}
    \begin{split}
        \boldsymbol{H}_{\text{up,Gauss}} &= \sum_{j=1}^{20} \left[ \tilde{q} \boldsymbol{N}^j + \frac{\tilde{c}_1}{2} \left( \boldsymbol{a}^j \boldsymbol{a}^j + \boldsymbol{a}^{j \dagger} \boldsymbol{a}^{j \dagger} \right) \right] \\
        &- J \sum_{j=1}^{19} \left( \boldsymbol{a}^{j \dagger} \boldsymbol{a}^{j + 1} + \boldsymbol{a}^{(j+1) \dagger} \boldsymbol{a}^j \right).
    \end{split}
    \label{eq:GaussianHamiltonian}
\end{equation}
Here, we introduced the relative mode operators $\boldsymbol{a}^j = ( \boldsymbol{a}_{1}^j + \boldsymbol{a}^j_{-1} ) / \sqrt{2}$ as well as the rescaled couplings $\tilde{c}_1 = c_1 n$ and $\tilde{q} = c_1 \left(n - \frac{1}{2} \right) + q$.

Beyond this regime, the high occupation justifies employing semi-classical approaches such as the truncated Wigner approximation (TWA), in which the mode operators are demoted to $c$-numbers that obey an evolution dictated by classical mean field equations \cite{Polkovnikov2010,Blakie2008}. The resulting model correctly captures the quantum fluctuations of the initial state while neglecting higher-order corrections in $\hbar$ for its evolution.

\textit{Measurement distributions} --- In what follows, we investigate the open-system dynamics of the five middle wells $8 - 12$, which we refer to as $AB$, see Fig.~\hyperref[fig:Intro]{1b)}. Given the locality of the interactions in both \eqref{eq:FullHamiltonian} and \eqref{eq:GaussianHamiltonian} and the product-form of the initial state, a dynamic build-up of an area law is expected \cite{Eisert2010}. 

We analyze the information content of the five-well system $AB$ in terms of measurement distributions using phase-space methods, see Fig.~\hyperref[fig:Intro]{1c)}. We focus on the two normalized spin-1 observables \cite{Hamley2012,Kawaguchi2012}
\begin{equation}
    \begin{split}
        \boldsymbol{\phi}^j &\equiv \frac{\boldsymbol{S}_x^j}{\sqrt{2n}} = \frac{1}{\sqrt{2}} \left[\boldsymbol{a}_0^{j \dagger} \left( \boldsymbol{a}^j_{1} + \boldsymbol{a}^j_{-1} \right) + h.c. \right]/\sqrt{2n},\\
        \boldsymbol{\pi}^j &\equiv -\frac{\boldsymbol{Q}_{yz}^j}{\sqrt{2n}} = \frac{-i}{\sqrt{2}} \left[\boldsymbol{a}_0^{j \dagger} \left( \boldsymbol{a}^j_{1} + \boldsymbol{a}^j_{-1} \right) - h.c.\right]/\sqrt{2n},
    \end{split}
    \label{eq:RelationSpinOperatorsConjugateOperators}
\end{equation}
which form a set of pairwise canonically conjugate operators $[\boldsymbol{\phi}^j, \boldsymbol{\pi}^{j'}] = i \delta^{j j'} \mathds{1}$ with corresponding bosonic mode operators $\boldsymbol{a}^{j}, \boldsymbol{a}^{j\dagger}$ in the early-time regime \cite{Haas2025}. 

Information about these observables is encoded in various measurement distributions $\mathcal{O}^j \equiv \mathcal{O}^j (\phi^j, \pi^j)$. One possibility is to consider the Wigner $W$-distribution defined via \cite{Weedbrook2012}
\begin{equation}
    \begin{split}
        \mathcal{W}^j (\phi^j, \pi^j) &= \int \frac{\mathrm{d} \tilde{\phi}^{j} \, \mathrm{d} \tilde{\pi}^{j}}{2 \pi} \, e^{-i (\phi^j, \pi^j) \Omega (\tilde{\phi}^j, \tilde{\pi}^j)^T} \\
        &\hspace{0.9cm} \times \Tr \left\{ \boldsymbol{\rho}^j \, e^{i (\boldsymbol{\phi}^j, \boldsymbol{\pi}^j) \Omega (\tilde{\phi}^j, \tilde{\pi}^j)^T} \right\},
    \end{split}
    \label{eq:WignerWDistribution}
\end{equation}
with the symplectic form $\Omega=i\sigma_2$ and $\sigma_2$ being the second Pauli matrix. As $\mathcal{W}^j$ is only accessible through costly Wigner tomography \cite{Schleich2001,Mandel2013}, it is mainly of theoretical interest. Thus, we also introduce experimentally more convenient distributions, namely the Wigner marginals $f^j(\phi^j) = \int \mathrm{d}\pi^j \mathcal{W}^j$ and $g^j (\pi^j) = \int \mathrm{d}\phi^j \mathcal{W}^j$, accessible through homodyne measurements \cite{Leonhardt1995}, as well as the Husimi $Q$-distribution, which is obtained by projecting onto the coherent states $\ket{\alpha^j} = \exp{(\alpha^j \boldsymbol{a}^{j \dagger} - \alpha^{j*} \boldsymbol{a}^j)} \ket{0^j}$ \cite{Strobel2014,Kunkel2018,Kunkel2019, Kunkel2019b,Kunkel2021, Husimi1940,Cartwright1976}, with $\alpha^j=(\phi^j+i\pi^j)/\sqrt{2}$, leading to
\begin{equation}
    \mathcal{Q}^j (\phi^j, \pi^j) = \Tr \left\{ \boldsymbol{\rho}^j \ket{\boldsymbol{\alpha}^j} \bra{\boldsymbol{\alpha}^j} \right\}.
    \label{eq:HusimiQDistribution}
\end{equation}

\textit{Information and correlations from classical distributions} --- To analyze the information content of subsystem $A$, we consider any of the outcome distributions $\mathcal{O}^A$ defined with respect to the local state $\boldsymbol{\rho}^A$ [see Fig.~\hyperref[fig:Intro]{1b)}]. We define their differential entropies as
\begin{equation}
    S (\mathcal{O}^A) = - \int \mathrm{d} \nu^A \, \mathcal{O}^A \, \ln \mathcal{O}^A,
    \label{eq:ClassicalEntropy}
\end{equation}
where the integral measure $\mathrm{d}\nu^A$ runs over all corresponding degrees of freedom in $A$ and hence depends on the distribution under scrutiny \footnote{We have $\mathrm{d}\nu^A = \mathrm{d} \phi^A \mathrm{d}\pi^A$ for $\mathcal{O}^A = \mathcal{W}^ A$, $\mathrm{d}\nu^A = \mathrm{d} \phi^A$ ($\mathrm{d}\nu^A = \mathrm{d} \pi^A$) for $\mathcal{O}^A = f^A$ ($\mathcal{O}^A = g^A$) and $\mathrm{d}\nu^A = \mathrm{d} \phi^A \pi^A / (2 \pi)^{\text{dim}\,A}$ for $\mathcal{O}^A = \mathcal{Q}^A$}.

We note that \eqref{eq:ClassicalEntropy} is always well-defined for the non-negative marginal and Husimi $Q$-distributions but is restricted to Wigner-positive states when applied to $\mathcal{W}^A$, which is an assumption implicitly made when working within TWA or Gaussian models. 

Being measures of disorder, classical entropies over incompatible observables are bounded from below by their vacuum values via entropic uncertainty relations \cite{Bialynicki-Birula1975,Wehrl1979,Lieb1978,VanHerstraeten2021a,Haas2021b,Haas2024} (see \cite{Coles2017,Hertz2019} for reviews). When considered for quantum many-body systems, the entropies of the local vacuum distributions $\bar{\mathcal{O}}^A$ scale with the number of modes, i.e., $S (\bar{\mathcal{O}}^A) \sim M$, showing that classical entropies are extensive to leading order as a result of vacuum contributions \cite{Haas2022b,Haas2023a}. However, as shown in \cite{Haas2023b}, scalings induced by quantum phenomena, such as the area law, manifest themselves in the \textit{next-to-leading} order terms. Thus, we define the so-called subtracted classical entropies as \cite{Haas2023b}
\begin{equation}
    \Delta S (\mathcal{O}^A) \equiv S (\mathcal{O}^A) - S (\bar{\mathcal{O}}^A),
    \label{eq:SubtractedClassicalEntropy}
\end{equation}
with the extensive vacuum contribution $S (\bar{\mathcal{O}}^A) \sim M$ being subtracted \footnote{The proportionality constant is $1 +  \ln \pi, (1+\ln \pi)/2$ and $1$ for the Wigner $W$, the marginal and the Husimi $Q$-distribution, respectively}.

Let us further consider the classical version of the archetypical measure for correlations between the left and right parts of the subsystem, that is, the classical mutual information
\begin{equation}
    I (\mathcal{O}^A : \mathcal{O}^B) = S(\mathcal{O}^A) + S(\mathcal{O}^B) - S(\mathcal{O}).
    \label{eq:ClassicalMutualInformation}
\end{equation}
Being already defined via a relative entropic measure, no vacuum contributions have to be subtracted to reveal quantum features.

\textit{Connections to quantum information theory} --- In the context of the Gaussian model \eqref{eq:GaussianHamiltonian}, the connection between subtracted classical and quantum entropies becomes a simple equality: in this case, we can establish $\Delta S (\mathcal{W}^A) = S_2 (\boldsymbol{\rho}^A)$, where $S_2(\boldsymbol{\rho}^A)$ denotes the Rényi-2 entropy of the density matrix associated to $\mathcal{W}^A$ \cite{Adesso2012}. Beyond Gaussianity, such simple relations can only be established for the subtracted Rényi-2 entropy of $\mathcal{W}^A$ \cite{Serafini2017}. However, in the following, we provide strong evidence that the scaling of the subtracted classical entropies \eqref{eq:SubtractedClassicalEntropy} also extends to the non-Gaussian interacting case.

Furthermore, a connection to the quantum mutual information in the case of Gaussian states is straightforward and reads $I (\mathcal{W}^A : \mathcal{W}^B) = I_2 (\boldsymbol{\rho}^A : \boldsymbol{\rho}^B)$ \cite{Adesso2012}. More generally, classical mutual informations constitute lower bounds to their quantum analogs by the uncertainty principle, i.e., \cite{Lieb2005,Haas2021b}
\begin{equation}
    I (\mathcal{O}^A : \mathcal{O}^B) \le I (\boldsymbol{\rho}^A : \boldsymbol{\rho}^B),
    \label{eq:EURMutualInformation}
\end{equation}
which are expected to be tighter than second-moment bounds beyond Gaussian states \footnote{The upper bound $I (f^A : f^B) + I (g^A : g^B) \le I (\boldsymbol{\rho}^A : \boldsymbol{\rho}^B)$ has been conjectured in \cite{Schneeloch2014}}. An immediate consequence of \eqref{eq:EURMutualInformation} is that the standard argument for the appearance of the area law for local interactions and thermal states presented in \cite{Wolf2008} also applies to any classical mutual information \cite{Haas2023b}. Hence, classical mutual information, albeit typically not capturing all quantum correlations, exhibits an area law whenever its quantum analog does. While the reverse statement does not follow from \eqref{eq:EURMutualInformation}, if one finds the classical mutual information to follow a volume law, the quantum one does as well. Note that for globally pure states, the same arguments apply to the entanglement entropy, in which case $I (\boldsymbol{\rho}^A : \boldsymbol{\rho}^A) = 2 S(\boldsymbol{\rho}^A)$.

\begin{figure*}[t!]
    \centering
    \includegraphics[width=.99\linewidth]{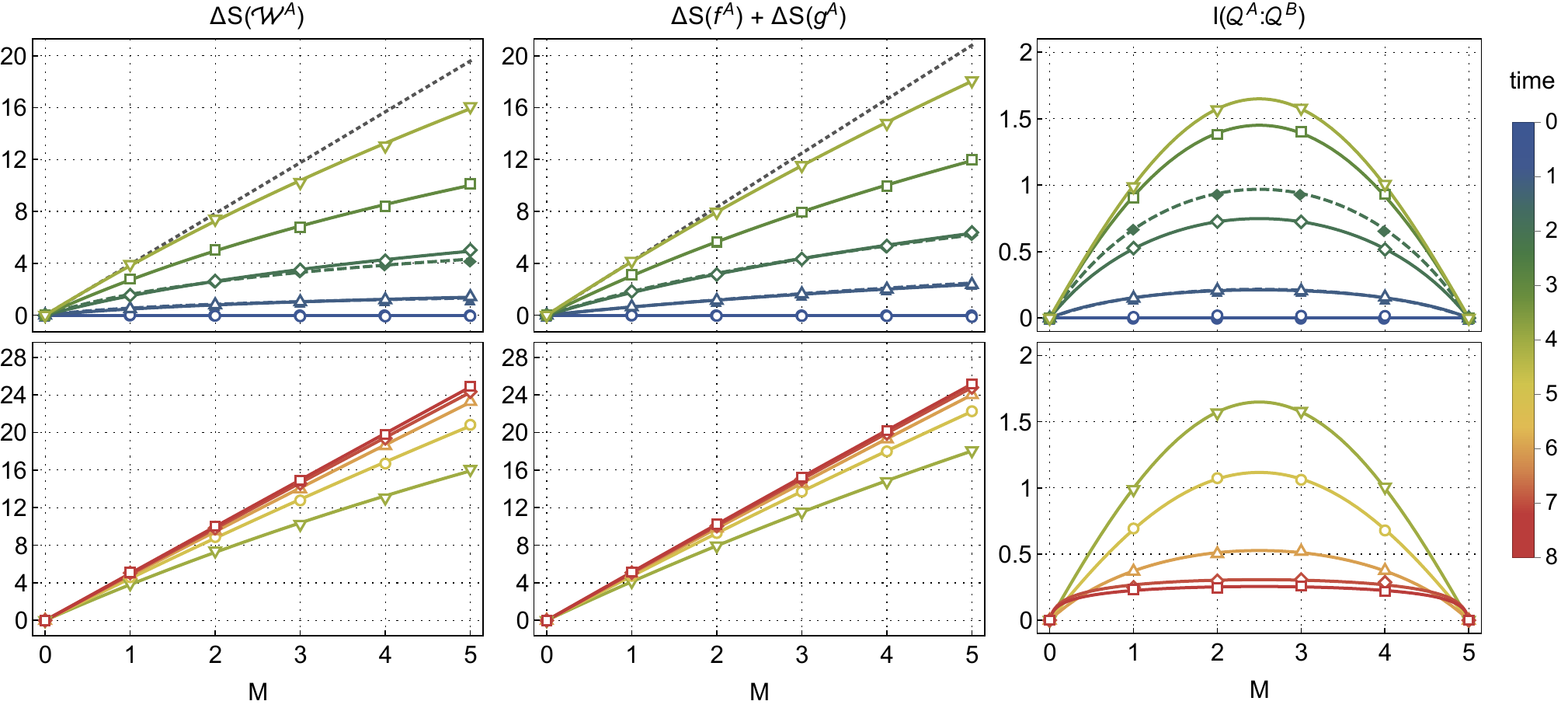}
    \caption{Analysis regarding the presence of area and volume laws at early times $t=0,1,2,3,4$ (upper row) and late times $t=4,5,6,7,8$ (lower row), respectively. Open (closed) plot markers denote TWA (analytic) results, and the corresponding solid (dashed) curves are fits. In the early-time regime, we observe the subtracted classical entropies to fulfill a logarithmic growth with subsystem size in the sense of \eqref{eq:AreaLaw} (see \cite{SM} for the \textit{standard} Wigner entropy). Their sublinear scaling is highlighted for $t=4$ by straight lines (gray dotted), which are fitted to the first two data points. In accordance, we also find the finite-size area law \eqref{eq:FiniteSizeAreaLaw} for the Wehrl mutual information. These findings hold for both the TWA and the analytical approach, which agree in the Gaussian regime, i.e., up to $t = 3$ \cite{SM}, thereby also validating the $k$NN estimator. For later times, the area law of the subtracted classical entropies tends to a stationary volume law \eqref{eq:VolumeLaw}, thereby demonstrating local thermalization. After the stationary point $t=7$, the local temperature can be extracted via their inclines, which consistently yields $T \approx 5$. The appearance of local thermalization is further supported by the decreasing correlations between $A$ and $B$ towards zero, as revealed by the evolution of the Wehrl mutual information.}
    \label{fig:Entropies}
\end{figure*}

\textit{Methods} --- We generate $10^4$ synthetic samples for the three distributions of our interest using TWA to simulate an experiment showcasing the feasibility of the proposed approach. In contrast to the estimation of low-order moments, extracting entropic quantities from a set of samples is more involved since they are \textit{functionals} of the underlying distributions. However, estimating an entropy from samples is still less demanding than reconstructing the underlying distribution. Given a set of samples, we employ the established $k$-nearest neighbor ($k$NN) method devised in \cite{Kozachenko1987, Kraskov2004, SteegGitHub} using information about the statistics of the nearest neighbors of each sample [see Fig.~\hyperref[fig:Intro]{1c)}], to arrive at an estimate of its local density. These results are validated against the analytically solvable model \eqref{eq:GaussianHamiltonian} in the early-time regime. We give a more comprehensive validation of the $k$NN-estimator for our setup in \cite{Haas2025}. 

We define an energy scale by setting $n c_1=-1$, which renders all quantities of interest dimensionless. We consider Lithium-7 with $c_0=-2c_1$, in which case $\abs{n c_1} \approx 100$Hz for $n = 1000$ atoms per well \cite{Huh2020}. Further, we set the quench parameters to $q=2J=4$, such that non-Gaussian features arise around $t = 3$, see \cite{SM} (see also \cite{Haas2025} for other parameter choices). Here, $t=1$ corresponds to one spin oscillation time, for which $t \approx 6$ms was reported in \cite{Huh2020}.

While the total system of 20 wells undergoes a unitary evolution dictated by the Hamiltonian in Eq.~\eqref{eq:FullHamiltonian}, the considered system $AB$ does not, as its entanglement with the rest of the system implies a mixed reduced density matrix \cite{Pechukas1994}. In the following, we demonstrate the area law and local thermalization for the theoretically interesting but experimentally difficult to access subtracted Wigner entropy, as well as for the experimentally amenable subtracted marginal entropy sum $\Delta S (f^A) + \Delta S (g^A)$, and the so-called Wehrl mutual information $I (\mathcal{Q}^A : \mathcal{Q}^B)$ (additional quantities are discussed in \cite{SM}).

\textit{Area law} --- We first study the early-time regime, that is, $t \le 4$, in the upper row of Fig.~\hyperref[fig:Intro]{2}. At $t=0$, the subsystem is in a pure product state, and all entropic measures evaluate to zero \footnote{See \cite{Haas2025} for a discussion on a thermal initial state}. Around $t=1$, correlations among the wells start to build up, causing subsystem $A$ to become entangled with its complement $B$. In this regime, subtracted classical entropies obey the area law, i.e., a logarithmic growth with system size $M$,
\begin{equation}
    \Delta S (\mathcal{O}^A) = \kappa_1 \ln \left( M + \kappa_2 \right) + \kappa_3,
    \label{eq:AreaLaw}
\end{equation}
just as one would expect for the entanglement entropy \cite{Calabrese2004,Calabrese2006,Plenio2007,Hastings2007,Amico2008,Calabrese2009,Casini2009,Peschel2009,Eisert2010}. The fit parameters $\kappa_i$ are constrained by $\kappa_2 = e^{- \kappa_3 / \kappa_1}$ to ensure $\Delta S(\mathcal{O}^A)=0$ when $M=0$. For $1.5 \lesssim t \lesssim 3$, a Bayesian hypothesis test with Gaussian noise mimicking experimental imperfections shows that the likelihood of a logarithmic scaling exceeds linear models, see \cite{SM}, thereby backing the area law's practical accessibility. Around $t = 3$, the distributions begin to exhibit non-Gaussian features, which we quantify by the relative entropy with respect to the closest Gaussian distribution, see \cite{SM}.

Similarly, the Wehrl mutual information signals the generation of correlations between $A$ and $B$ in terms of the finite-size area law \cite{Calabrese2004}
\begin{equation}
    I(\mathcal{O}^A:\mathcal{O}^B) = \kappa_1 \ln \left[ \frac{5}{\pi} \sin \left(\frac{\pi M}{5} \right) + \kappa_2 \right] + \kappa_3,
    \label{eq:FiniteSizeAreaLaw}
\end{equation}
which incorporates the reflection symmetry around $M=2.5$. Again, the behavior coincides with what is expected for the quantum mutual information \cite{Wolf2008}, with maximal correlations occurring at $t=4$.

\textit{Local thermalization} --- For later times, i.e., in the regime $t \ge 4$ (lower row of Fig.~\hyperref[fig:Intro]{2}), the subtracted classical entropies transition from an intermediate stage around $t = 5$ to an extensive growth with system size at $t=7$. The latter remains stationary beyond $t=7$, signaling that the system has thermalized locally in the considered degrees of freedom, with the remaining system serving as a heat bath. In this case, all entropies of our interest obey the volume law \cite{Abanin2019}
\begin{equation}
    \Delta S (\mathcal{O}^A) = \beta \, M,
    \label{eq:VolumeLaw}
\end{equation}
where $\beta = 1/T$ denotes the inverse local temperature. Indeed, both final entropic curves show an incline of $T \approx 5$, illustrating how the local temperature can be extracted from classical entropies by simple means. We have checked that this temperature depends only weakly on the quench parameters, as the dominating energy scale is set by the fourth-order term proportional to $c_0$ in \eqref{eq:FullHamiltonian}.

While the classical entropies become extensive, the Wehrl mutual information still obeys the finite-size area law \eqref{eq:FiniteSizeAreaLaw} for later times, which also highlights its robustness against thermal fluctuations. In contrast to the early-time dynamics, the correlations between $A$ and $B$ now decline monotonically towards local thermal equilibrium.

\textit{Discussion} --- 
We have demonstrated that quantum many-body phenomena could be probed with classical entropies by considering a concrete model system that can be readily realized experimentally. Specifically, we have shown that it is possible to observe the area law, that is, the characteristic logarithmic growth of the entanglement entropy, and the volume law, which indicates local thermalization, via subtracted classical entropies and mutual informations of experimentally accessible measurement distributions. Crucially, we have not assumed the state to obey a specific functional form. We bypassed the similarly costly reconstruction of a measurement distribution by estimating its classical entropy directly from the sampled data. In this way, we relied on $10^4$ samples only -- even in the strongly non-Gaussian regime and up to ten-dimensional distributions -- which we deem experimentally feasible. Larger system sizes ($M>5$) can be tackled with comparable sample numbers provided that the sampled data still captures essential features of the underlying distribution. Future work will address what other parallels between classical entropies and quantum entropies exist, especially for other degrees of freedom, and whether they also lend themselves as easily to experimental implementations as in the discussed work.

\textit{Acknowledgements} --- The authors thank Thomas Gasenzer and Ido Siovitz for helpful discussions during the development of the manuscript and an anonymous referee for valuable suggestions. T. H. is supported by the European Union under project ShoQC within the ERA-NET Cofund in Quantum Technologies (QuantERA) program, as well as by the F.R.S.- FNRS under project CHEQS within the Excellence of Science (EOS) program. This work is supported by the Deutsche Forschungsgemeinschaft (DFG, German Research Foundation) under Germany’s Excellence Strategy EXC2181/1-390900948 (the Heidelberg STRUCTURES Excellence Cluster) and within the Collaborative Research Center SFB1225 (ISOQUANT). This work was partially financed by the Baden-Württemberg Stiftung gGmbH. The authors gratefully acknowledge the Gauss Centre for Supercomputing e.V. (www.gauss-centre.eu) for funding this project by providing computing time through the John von Neumann Institute for Computing (NIC) on the GCS Supercomputer JUWELS \cite{JUWELS} at Jülich Supercomputing Centre (JSC).


\bibliography{references.bib}

\end{document}


\title{Supplementary material}

\maketitle

\section{Classical Wigner entropy}
\begin{figure}[h!]
    \centering
    \includegraphics[width=0.95\linewidth]{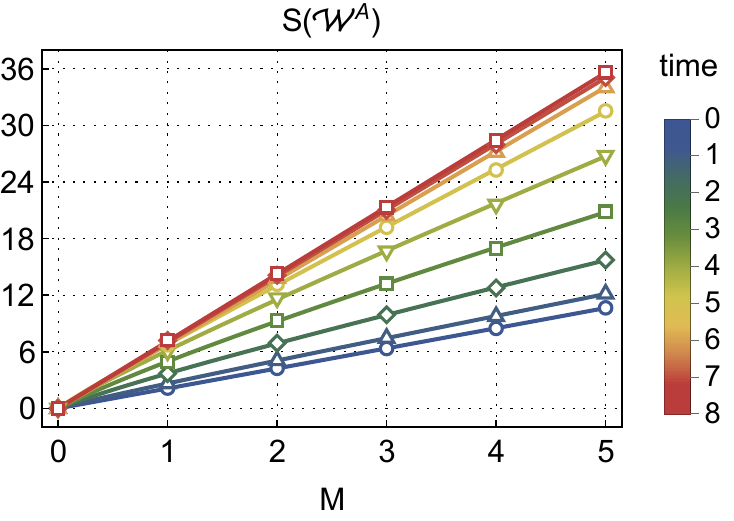}
    \caption{Time evolution of the full Wigner entropy $S(\mathcal{W}^A)$ for which the leading-order volume law is apparent for all times. The area law is barely visible on top of the extensive growth in the early-time regime, i.e., for $0 < t \le 4$. Note that at $t=0$ we have $S(\mathcal{W}^A) = S(\bar{\mathcal{W}}^A) = M (1 + \ln \pi) \approx 2.144 M$.}
    \label{fig:WignerEntropy}
\end{figure}

We illustrate the extensive growth of standard classical entropies by plotting the full classical Wigner entropy, i.e., without subtracting the vacuum contribution, in Fig.~\hyperref[fig:WignerEntropy]{1}.

\section{Non-Gaussianity}
We consider a Gaussian model distribution 
\begin{equation}
    \mathcal{W}^{A,\text{Gauss}} = \frac{1}{Z^A} \, e^{- \frac{1}{2} (\chi^A)^T (\gamma^A)^{-1} \chi^A},
    \label{eq:GaussianWignerWDistribution}
\end{equation}
where $\chi^A = (\phi^A, \pi^A)^T$ is a vector in phase space, $(\gamma^A)^{j j'} = \Tr \{ \boldsymbol{\rho}^A \{ \boldsymbol{\chi}^j - \chi^j, \boldsymbol{\chi}^{j'} - \chi^{j'} \} \}/2$ denotes the covariance matrix and $Z^A = (2 \pi)^M \, \sqrt{\det \gamma^A}$ is a normalization constant. To assess the non-Gaussianity of a given distribution $\mathcal{W}^A$, we introduce the Wigner relative entropy with respect to the nearest Gaussian, i.e., the distribution with the same first- and second-order moments \cite{Haas2022b,Haas2023a}
\begin{equation}
    S (\mathcal{W}^A \| \mathcal{W}^{A,\text{Gauss}}) = \int \mathrm{d} \nu^A \, \mathcal{W}^A \, \ln \frac{\mathcal{W}^A}{ \mathcal{W}^{A,\text{Gauss}}}.
    \label{eq:WignerRelativeEntropy}
\end{equation}
Then, $\mathcal{W}^A$ is (non-)Gaussian if and only if $S (\mathcal{W}^A \| \mathcal{W}^{A,\text{Gauss}}) (>)=0$. The non-negativity of the Wigner relative entropy translates into a Gaussian upper bound on the subtracted Wigner entropy, i.e., $\Delta S (\mathcal{W}^A) \le \Delta S (\mathcal{W}^{A,\text{Gauss}})$, showing that resolving non-Gaussian features decreases the missing information about the underlying distribution. In this sense, $S (\mathcal{W}^A \| \mathcal{W}^{A,\text{Gauss}})$ measures the additional information encoded in $\mathcal{W}^A$ with respect to $\mathcal{W}^{A,\text{Gauss}}$.

\begin{figure}[h!]
    \centering
    \includegraphics[width=0.95\linewidth]{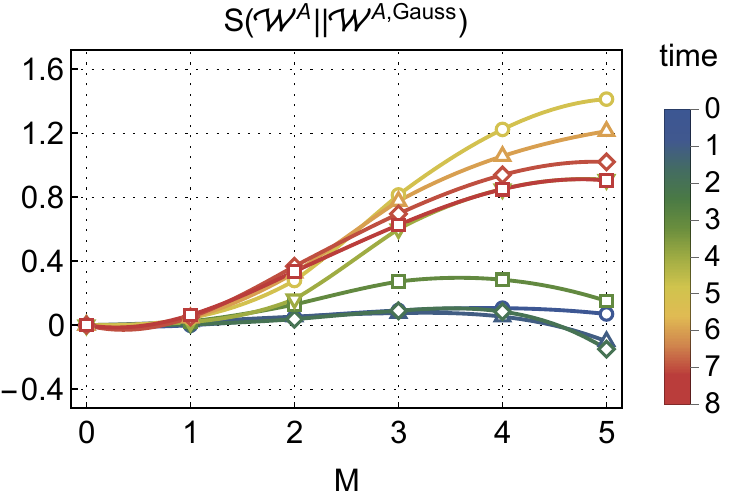}
    \caption{Time evolution of the non-Gaussianity measure $S (\mathcal{W}^A \| \mathcal{W}^{A,\text{Gauss}})$. Single-well distributions look rather Gaussian, while non-Gaussian features become apparent for larger subsystems. The non-Gaussianity is negligible for early times and peaks around $t \approx 5$, for which the relative information difference is $\sim 8 \%$. We checked that negative values at $M=5$ for early times are caused by an insufficient number of samples; see \cite{Haas2025} for details.}
    \label{fig:RelativeEntropies}
\end{figure}

To calculate the Wigner relative entropy \eqref{eq:WignerRelativeEntropy} without reconstructing any distribution, we use \eqref{eq:GaussianWignerWDistribution} and perform a few straightforward simplifications, leading to
\begin{equation}
    S (\mathcal{W}^A \| \mathcal{W}^{A,\text{Gauss}}) = \Delta S (\mathcal{W}^{A,\text{Gauss}}) - \Delta S (\mathcal{W}^A).
    \label{eq:WignerRelativeEntropyGaussian}
\end{equation}
While $\Delta S (\mathcal{W}^A)$ is estimated using the $k$NN method, the subtracted Wigner entropy of the nearest Gaussian distribution is computed via
\begin{equation}
    \Delta S (\mathcal{W}^{A,\text{Gauss}}) = \frac{1}{2} \ln \det \left( 2 \gamma^A \right),
\end{equation}
such that only the covariance matrix has to be extracted from the TWA samples. We show the resulting relative entropy curves in Fig.~\hyperref[fig:RelativeEntropies]{2} for all times discussed in the main text.

\begin{figure*}[t!]
    \centering
    \includegraphics[width=0.99\linewidth]{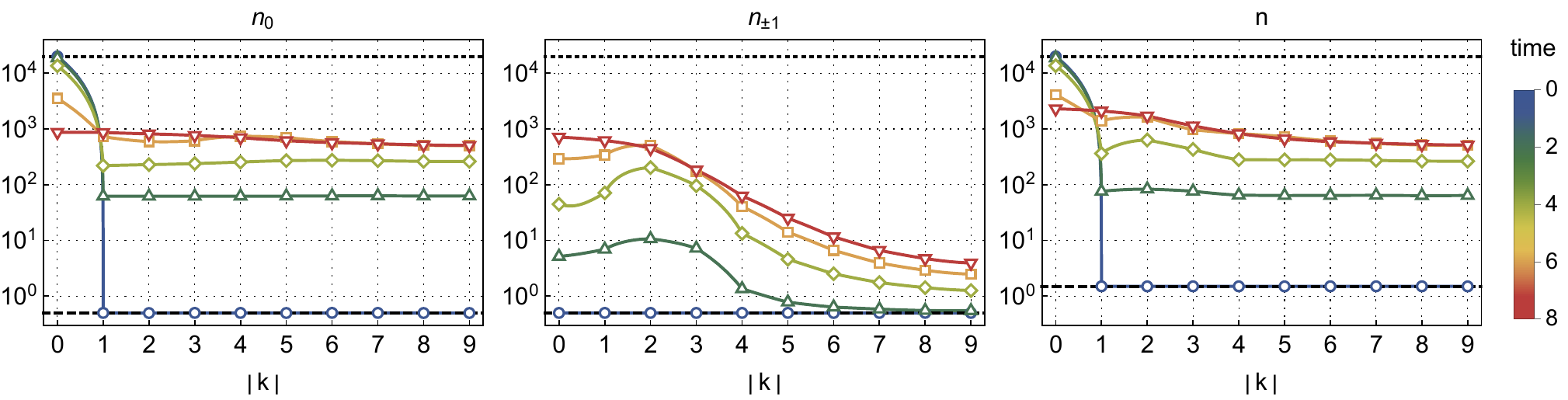}
    \caption{Dynamics of the momentum-mode occupations for the zero mode (left), the side modes (middle), and their sum (right). The atom number $n=2 \times 10^4$ and the quantum one-half (three-half) are depicted by dotted and dashed lines, respectively.}
    \label{fig:ModePopulation}
\end{figure*}

\section{Mode occupations for late times}
\textit{A priori}, it is unclear whether TWA gives meaningful results in the late-time limit where local thermalization occurs. As a semi-classical approximation, TWA is expected to hold whenever the momentum modes are occupied mesoscopically, that is, filled up to at least roughly one order of magnitude above the quantum one-half \cite{Werner1997, Steel1998, Sinatra2001, Sinatra2002}. In Fig.~\hyperref[fig:ModePopulation]{3}, we confirm that this condition is fulfilled for late times by plotting the momentum-mode occupations $n_{m_F} (k) = \braket{\mathbf{a}_{m_F}^{k \dagger} \mathbf{a}_{m_F}^{k}}$ for the zero mode (left panel), the side modes (middle panel) and their sum (right panel).

\begin{figure*}[t!]
    \centering
    \includegraphics[width=0.99\linewidth]{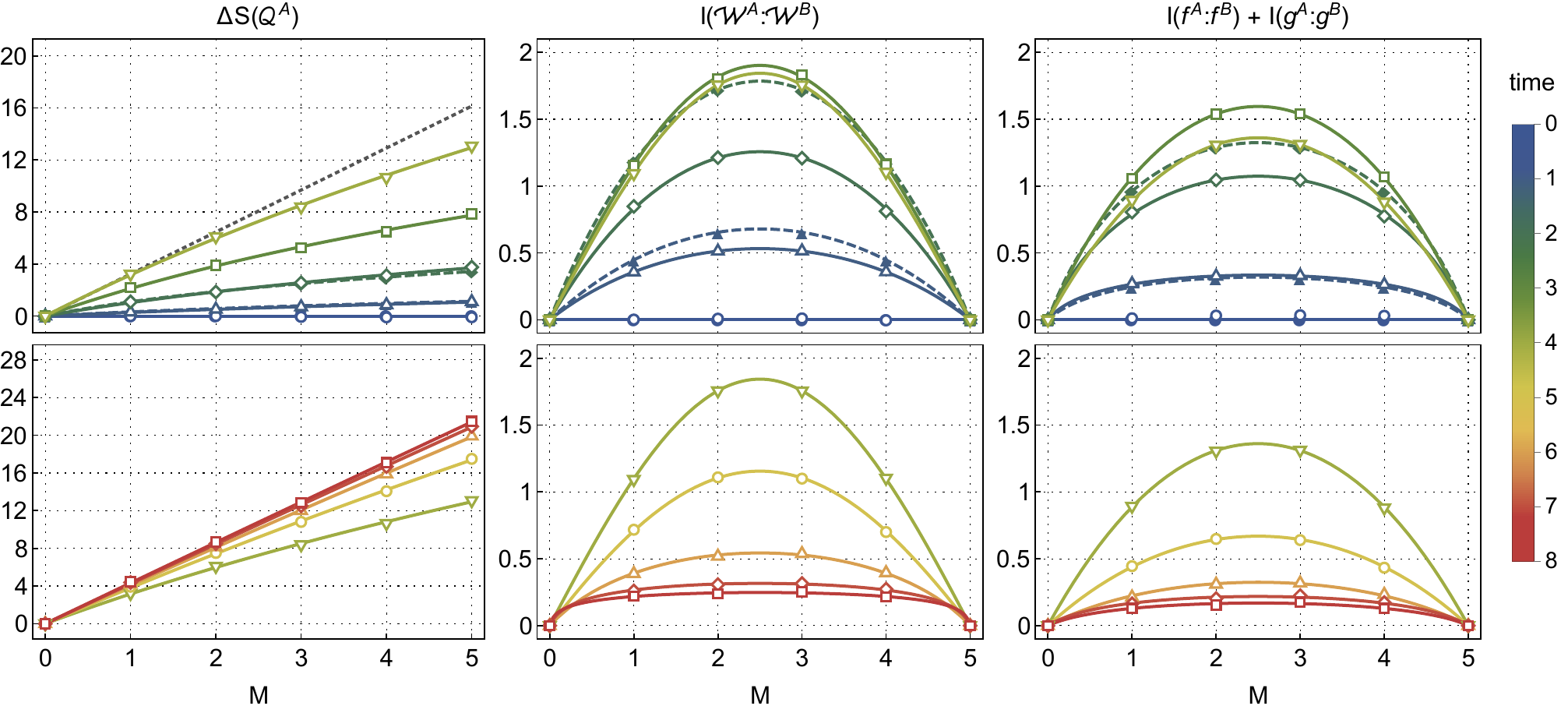}
    \caption{Same analysis as in Fig. 2 in the main text for the subtracted Wehrl entropy (left column), the Wigner mutual information (middle column), and the marginal mutual information sum (right column). All observed quantum features carry over to these three quantities as well. The local temperature $T \approx 5$ is also observed for the subtracted Wehrl entropy. Note here that the latter is based on the differently normalized Husimi $Q$-distribution, which we accounted for by subtracting $M \ln 2$.}
    \label{fig:EntropiesSM}
\end{figure*}

\section{Other classical information-theoretic measures}
In analogy to Figure 2 in the main text, we show the dynamics of the subtracted Wehrl entropy $\Delta S (\mathcal{Q}^A)$, the Wigner mutual information $I (\mathcal{W}^A : \mathcal{W}^B)$ and the marginal mutual information sum $I (f^A : f^B) + I (g^A : g^B)$ in Fig.~\hyperref[fig:EntropiesSM]{4}. All quantities behave as expected.

\begin{figure*}[t!]
    \centering
    \includegraphics[width=0.99\linewidth]{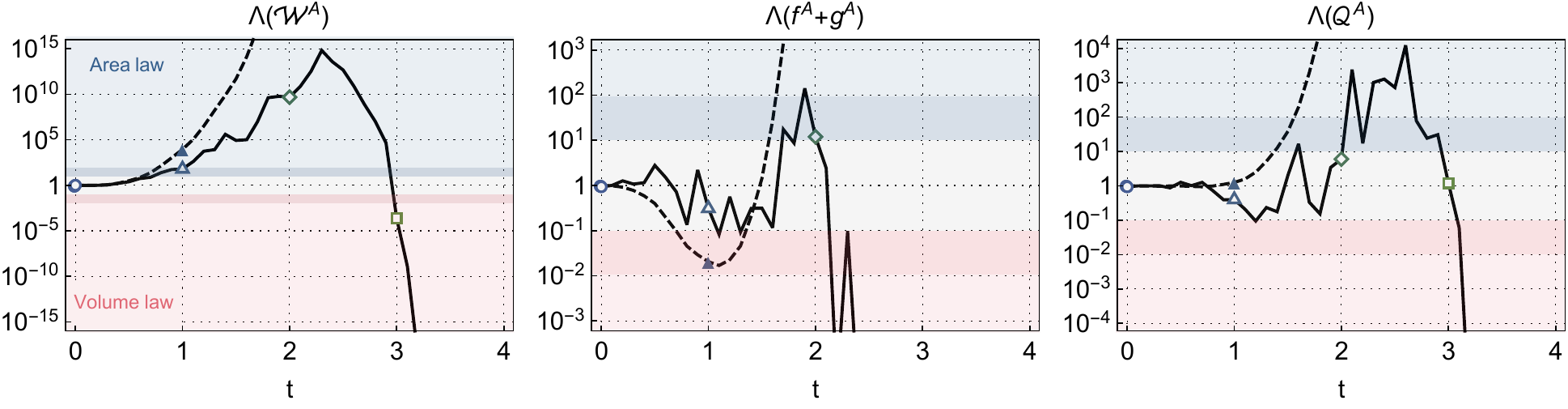}
    \caption{Ratios of the likelihoods associated with logarithmic (area-law-like) and linear (volume-law-like) scaling for the three subtracted classical entropies as a function of time (integer-valued times considered in the main text are highlighted by their corresponding plot markers). In a Bayesian sense, evidence for the null hypothesis (area law fits better than volume law) is strong (decisive) when this ratio exceeds $10 \, (100)$ (blue regions).}
    \label{fig:Inference}
\end{figure*}

\section{Bayesian hypothesis test for area vs. volume law}
In any real experiment, the measured data is subjected to various kinds of noise. Given that the area law signal in the subtracted classical entropies is contained in the next-to-leading-order contributions, an analysis of its statistical likelihood over a standard linear fit is warranted. To this end, we perform the Bayesian maximum likelihood test by considering a Gaussian noise model for the subtracted entropy centered around the two models of our interest $\mathcal{M}_1 (\kappa; M) = \kappa_1 \ln (M + e^{- \kappa_2 / \kappa_1}) + \kappa_2$ and $\mathcal{M}_2 (\kappa; M) = \kappa_1 M$. This defines the corresponding likelihood functions
\begin{equation}
    \mathcal{L}_j (\mathcal{O}^A; \kappa) = \prod_{M=0}^{5} \frac{1}{\sqrt{2 \pi \sigma^2}} \, e^{- \frac{1}{2 \sigma^2} \left[\Delta S (\mathcal{O}^A; M) - \mathcal{M}_j (\kappa; M) \right]^2},
    \label{eq:Likelihood}
\end{equation}
where $\sigma$ denotes the standard deviation of the noise model. Motivated by the fluctuations already present in the subtracted entropies extracted from the TWA data at $t=0$ (where all entropies should be strictly zero), we estimate $\sigma \approx 0.1$, which roughly corresponds to twice the observed variations. To compare the two models, we compute the ratio of the likelihoods \eqref{eq:Likelihood} maximized over their model parameters
\begin{equation}
    \Lambda (\mathcal{O}^A) = \frac{\max_{\kappa_1, \kappa_2} \mathcal{L}_1 (\mathcal{O}^A; \kappa)}{\max_{\kappa_1} \mathcal{L}_2 (\mathcal{O}^A; \kappa)}.
\end{equation}

We show the resulting ratios for the three types of measurement distributions of our interest as a function of time in Fig.~\hyperref[fig:Inference]{5}. According to Jeffreys' scale, which constitutes a standard Bayesian decision criterion, the evidence for the first model (an area-law-fit) being more likely than the second (a volume-law-fit) is substantial (decisive) when $\Lambda > 10 \, (100)$, see blue regions, and vice versa if $\Lambda < 1/10 \, (1/100)$, see red regions. After some initial period where both models are equally likely, all subtracted entropies exhibit clear indications of logarithmic behavior. The evidence is most substantial for the subtracted Wigner entropy, for which the maximum likelihood ratio is well above $100$ for $t \lesssim 3$. For later times, the likelihood of linear behavior increases rapidly, thereby ruling out the area law in this regime.

\bibliography{references.bib}